\documentclass[sigconf]{acmart}
\AtBeginDocument{%
  \providecommand\BibTeX{{%
    \normalfont B\kern-0.5em{\scshape i\kern-0.25em b}\kern-0.8em\TeX}}}

\setcopyright{acmcopyright}

\copyrightyear{2023}
\acmYear{2023}
\setcopyright{rightsretained}
\acmConference[UbiComp/ISWC '23]{Proceedings of the 2023 ACM International Joint Conference on Pervasive and Ubiquitous Computing and the 2023 ACM International Symposium on Wearable Computers}{8-12
October, 2023}{Cancún, Mexico}
\acmBooktitle{roceedings of the ACM on Interactive, Mobile, Wearable and Ubiquitous Technologies (IMWUT), 8-12
October, 2023, Cancún, Mexico.}
\acmDOI{10.1145/3544793}
\acmISBN{978-1-4503-9423-9}

\begin{document}

\title{Inspire creativity with ORIBA: Transform Artists' Original Characters into Chatbots through Large Language Model}


\author{Yuqian Sun}
\authornotemark[1]
\affiliation{%
  \institution{Royal College of Art}
  \city{London}
  \country{United Kingdom}}
\email{yuqiansun@network.rca.ac.uk}
\author{Xinyu Li}
\authornotemark[1]
\affiliation{%
  \institution{Georgia Institute of Technology}
  \city{Atlanta}
  \country{United States}}
\email{xingyu@gatech.edu}
\author{Ze Gao}
\authornote{Both authors contributed equally to this research.}
\affiliation{%
  \institution{Hong Kong University of Science and Technology}
  \city{Hong Kong SAR}
  \country{China}}
\email{zgaoap@connect.ust.hk}

\renewcommand{\shortauthors}{Sun, Li and Gao.}

\begin{abstract}
This research delves into the intersection of illustration art and artificial intelligence (AI), focusing on how illustrators engage with AI agents that embody their original characters (OCs). We introduce 'ORIBA', a customizable AI chatbot that enables illustrators to converse with their OCs. This approach allows artists to not only receive responses from their OCs but also to observe their inner monologues and behavior. Despite the existing tension between artists and AI, our study explores innovative collaboration methods that are inspiring to illustrators. By examining the impact of AI on the creative process and the boundaries of authorship, we aim to enhance human-AI interactions in creative fields, with potential applications extending beyond illustration to interactive storytelling and more.
\vspace{-0.2cm}
\end{abstract}




\begin{teaserfigure}
    \vspace{-0.4cm}
  \includegraphics[width=\textwidth]{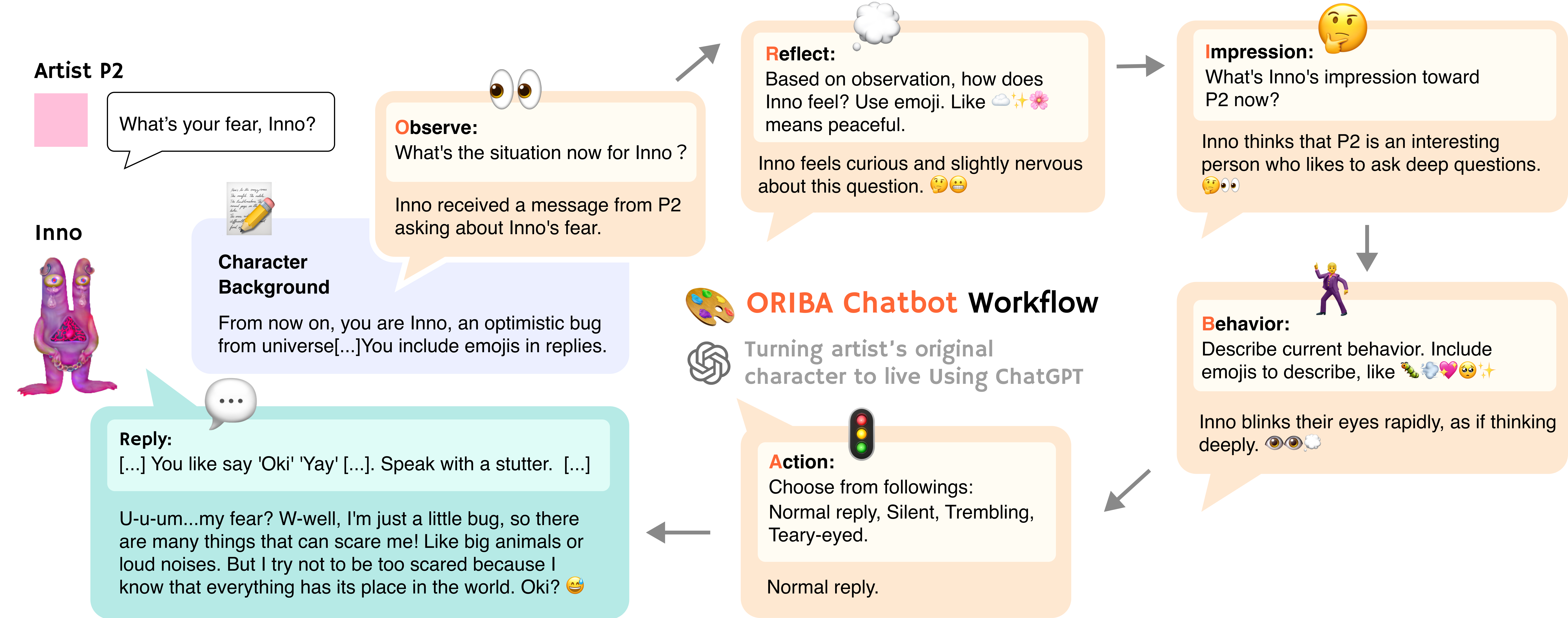}
  \caption{Artist P2's original character(OC) 'Inno' chatbot. The conversation is generated through ORIBA workflow: Observe, Reflect, Impression, Behavior, Action. This leads large language model(LLM) to generate in-depth and informative conversations.}
  \Description{Artist P2's OC 'Inno' and their conversations. P2 hopes Inno to express together with emojis, Artist P2's original character(OC) 'Inno' chatbot. The conversation is generated through ORIBA workflow: Observe, Reflect, Impression, Behavior, Action. This leads large language models (LLM) to generate in-depth and informative conversations, which will inspire artists in character creation.}
  \label{fig:teaser}
\end{teaserfigure}

\maketitle

\section{Introduction}
Imagine a world where artists can converse with the characters they create. This research explores this intriguing possibility, investigating how illustrators perceive the agency of Artificial Intelligence(AI) agents, particularly in the context of their original characters (OCs). Recent advancements in Large Language Models (LLMs) have demonstrated their ability to follow intricate instructions and generate coherent, creative text \cite{baktash2023gpt4}. These developments underscore the potential of LLMs to augment the believability of virtual characters. However, the human perception formed through interactions with these characters remains largely uncharted territory.

Our study delves into this fascinating domain, aiming to explore the creative process by engaging illustrators in a novel interaction with AI agents that represent their own OCs. We employ artificial intelligence technology to transform original characters, previously confined to visual representation, into interactive AI agents. This approach transforms the traditional "think-draw-publish" workflow, turning the characters created by artists into communicative entities. These characters can engage in discussions with the artists and invite the audience to participate in the narrative in which the character resides.

We have developed a customizable AI chatbot, 'ORIBA', which enables authors to engage in dialogue with their OCs through AI. 'ORIBA' is an acronym representing its workflow to configure the prompt to drive LLM ChatGPT\cite{ChatGPT}: Observe, Reflect, Impression (towards speaker), Behavior, and Action (Fig\ref{fig:teaser}). This approach allows artists to not only receive responses from their OCs but also observe how they think and behave. Through in-depth and informative conversations, we aim to support and facilitate the creative process of authors, providing an experimental platform for authors to verify and expand upon their creations.
\vspace{-0.2cm}
\section{Motivations and Contributions}

With the development of AI image generation technology, AI art generation tools like Midjourney\cite{midjourney} and Stable Diffusion\cite{stablediffusion} have begun to great attention\cite{Mind-Boggling}. However, the relationship between artists and AI has become increasingly tense. Artists have responded to these challenges through a variety of means, including litigation\cite{Against}, online boycotts, and petitions\cite{Artists}\cite{Creatives}. Three potential reasons may have contributed to great AI negative impact on illustrators: 1) Current policies may be insufficient in protecting the copyrights and intellectual property rights of original illustrators \cite{shanGLAZEProtectingArtists2023}. 2) The proliferation of AI-generated images damages the creative enthusiasm of artists and undermines their ability to make a living \cite{shanGLAZEProtectingArtists2023}. 3) Algorithm aversion, which posits that beautiful creations should be the work of humans, not AI \cite{10.1145/3349537.3351891}.

Despite these challenges, the widespread adoption of AI is an inevitable trend. Thus, we aim to explore new ways of collaborating with AI that can be accepted by illustrators. We believe that conversational AI can help illustrators address these challenges and incorporate AI agents' agency into their creative process.

Illustrators, whether individual creators or experts in companies, can benefit from a more vivid character experience. By transforming their original characters (OCs) into chatbots, illustrators can tell stories in a novel way. This not only provides a unique interaction for the audience but also gives illustrators direct feedback, which can be a source of encouragement for further creative work.

The emerging intelligence of large language models (LLMs) brings new potential to the creative process\cite{wei2022emergent}. The ability of AI agents to make autonomous decisions or mimic thought processes, such as the "Chain of Thoughts"\cite{wei2023chainofthought} method, has not yet been applied in Human-Computer Interaction (HCI). We believe that the expressive capabilities of these AI agents can enhance the creativity of illustrators.

Explore new ways for illustrators to collaborate with AI, instead of using image generation AI, collaborating with conversational AI agents involves a certain level of storytelling, including character backgrounds and dialogues, which can be expressed through text. This allows illustrators to interact directly with the characters they create. When these characters begin to exhibit agency, it raises questions about the boundaries of authorship. This could potentially blur the lines between creation and reality, sparking a wider discussion.

Although this research focuses on the illustration community, the technology used in this study, the transformation of original characters into interactive virtual entities through LLMs, has potential applications beyond illustration, especially in interactive AI literacy. The novel collaboration between the artist and characters will bring more engaging and dynamic human-AI interactions, including video games and creative education.

We construct a customizable chatbot 'ORIBA Agent' through LLM, enabling creators to engage in in-depth conversations with their OCs, reflecting on their inner monologues behind the replies. We aim to explore three main questions:1) Explore the collaborative experience between illustrators and AI in co-creation. Does the system inspire creators and have a positive impact on their enthusiasm and inspiration? 2) Study the influence of AI on the creative process. How our system affects the creative process and what's the impact of AI feedback on creators? 3) Discuss the boundaries of authorship in our system. How can we reconcile the authorship of the creator with AI-generated content?

\section{Related Works}

We reviewed three main points related to the use of artificial intelligence (AI) in creative tasks. Firstly, it discusses narrative intelligence and creating believable characters using AI. Secondly, it explores the use of autonomous agents for creators in generating content. Finally, it considers the potential of a creative AI assistant in supporting human creativity. It is clear that AI can be a valuable asset in creative tasks, despite its limitations and potential biases. Additionally, collaboration between creators and AI agents can lead to more efficient and productive results.

\subsection{Narrative Intelligence and believable character}

Narrative intelligence involves the ability to craft, tell, understand, and respond effectively to stories. Computational narrative intelligence aims to instill narrative intelligence into computers, making them better communicators, educators, and entertainers\cite{yannakakisPanoramaArtificialComputational2015}. In this context, AI agents with narrative intelligence can potentially become "living" social actors, engaging with human participants in a shared context.
Character believability on the other hand is the perception by the audience that the actions performed by characters do not negatively impact the audience's suspension of disbelief\cite{bogdanovychWhatMakesVirtual2016a}. Therefore, the implication of this is that if a character is perceived as believable; one should be able to infer its motivations and intentions through observations of the character. ORIBA workflow supports informative disclosure of character's inner monologue to make them believable.

\subsection{Autonomous Agents for creators}

The advent of Language Models (LLMs) as autonomous agents have given rise to projects such as AutoGPT\cite{AutoGPT2023}, BabyAGI\cite{babyAGI}, CAMEL\cite{Camel-AI2023}, and Generative Agents. These initiatives leverage LLMs as reasoning engines, equipped with tools for interfacing with data sources or computations, and memory for recalling past interactions.

The LangChain Agent\cite{langchain_2023}, based on the Reasoning and Acting (ReAct) framework\cite{yao2023react}, exemplifies this approach. The process involves a user task, agent thought, action decision, tool observation, and repetition until task completion.

Projects like AutoGPT\cite{AutoGPT2023} and BabyAGI\cite{babyAGI} emphasize long-term objectives and planning, applicable to artists' OCs for observation, reflection, and decision-making. Conversely, CAMEL and Generative\cite{park2023generative} Agents employ unique simulation environments and adaptive long-term memory.

Incorporating autonomous agents into the creative process allows for an exploration of their impact on artists' authorship. Their ability to plan, observe, and reflect could enhance artists' creative possibilities and provide new ways of interacting with their OCs. This could make characters more lifelike and potentially reshape artists' storytelling and illustration approaches.

By studying artists' perceptions of their OCs' autonomous agency, we can better understand the human-AI relationship in a creative context and contribute to the discourse on AI's role in arts and entertainment.
\vspace{-0.2cm}
\subsection{Creative AI assistant}

The literature on writing\cite{characterChat}\cite{scienceWriting} and drawing assistants\cite{openai2023dalle2} has grown significantly with the advancement of AI technology. Writing assistants, such as Grammarly and Hemingway, use Natural Language Processing (NLP) algorithms to provide suggestions for improving the clarity and readability of text. Drawing assistants, on the other hand, use machine learning to help artists create illustrations and design more efficiently. Examples of drawing assistants include Adobe's Sensei\cite{adobe2023sensei} and Google's AutoDraw\cite{google2017autodraw}. These tools aim to reduce the cognitive load on creators and improve the quality of their work. While some studies have shown that these assistants can significantly improve productivity and creativity, others have raised concerns about the potential loss of originality and creative control. Future research in this area will continue to explore the benefits and drawbacks of writing and drawing assistants, as well as their impact on the creative process.
\vspace{-0.3cm}
\section{Experiment Design}


Our experimental design serves as a blueprint for the scientific experiment, outlining the methods, variables, and participants involved to guarantee dependable and accurate outcomes. The methodology describes the step-by-step procedures, techniques, and tools employed during the research process, ensuring the study is replicable and reliable. The variables encompass the factors under investigation, including independent variables manipulated by the researcher and dependent variables measured in response to the independent variables. To control for potential confounding variables and minimize biases, researchers may identify and address extraneous variables that could influence the outcomes. Finally, the selection and recruitment of participants in the experiment are crucial to establishing external validity or the generalizability of the findings to a broader population. By carefully considering these components, an excellent experimental design enables researchers to draw meaningful conclusions from their observations and advance scientific understanding.

\vspace{-0.3cm}

\subsection{Participants}
We plan to recruit 15-20 illustrators to join the experiment. We published recruitment advertisements on the public social media platform, Weibo, and We request them to provide detailed information about the OC they are working on. Finally, we selected 4 illustrators (Table \ref{table: participants information}) as participants for a pre-study. The participants were not previously known to us. With an average age of 23, they had each been engaged in OC creation for more than three years. All participants are experienced in creating OCs. Additionally, all participants can express their feelings and opinions. 
\begin{table*}[t]
\vspace{-0.3cm}
\centering
\caption{Participants' and their OCs' information}
\vspace{-0.2cm}
\resizebox{1\textwidth}{!}{%
\small
\begin{tabular}{cccccccc}
\textbf{Participants} & \textbf{Age} & \textbf{Gender} & \textbf{Country} & \textbf{Years of Experience} & \textbf{OC Name} & \textbf{Character Language}  & \textbf{Occupation}                             \\ \hline
P1                    & 21           & F          & China            & 5 years                      & Unta (Deer Centaur)             & Chinese            & Animation                                       \\
P2                    & 29           & F          & China            & 7 years                      & Inno (Bug)            & English            & Industrial Design                               \\
P3                    & 23           & F          & China            & 13 years                     & Esca (Alien)            & Chinese and Non-human Language & Digital Media                                   \\
P4                    & 21           & F          & China            & 3 years                      & Devin (Human Soldier)        & Chinese    & Economic and Social History
\end{tabular}
}
\label{table: participants information}
\vspace{-0.3cm}
\end{table*}


\subsection{Experiment}


\begin{figure}[h]
\vspace{-0.2cm}
  \centering
  \includegraphics[width=\linewidth]{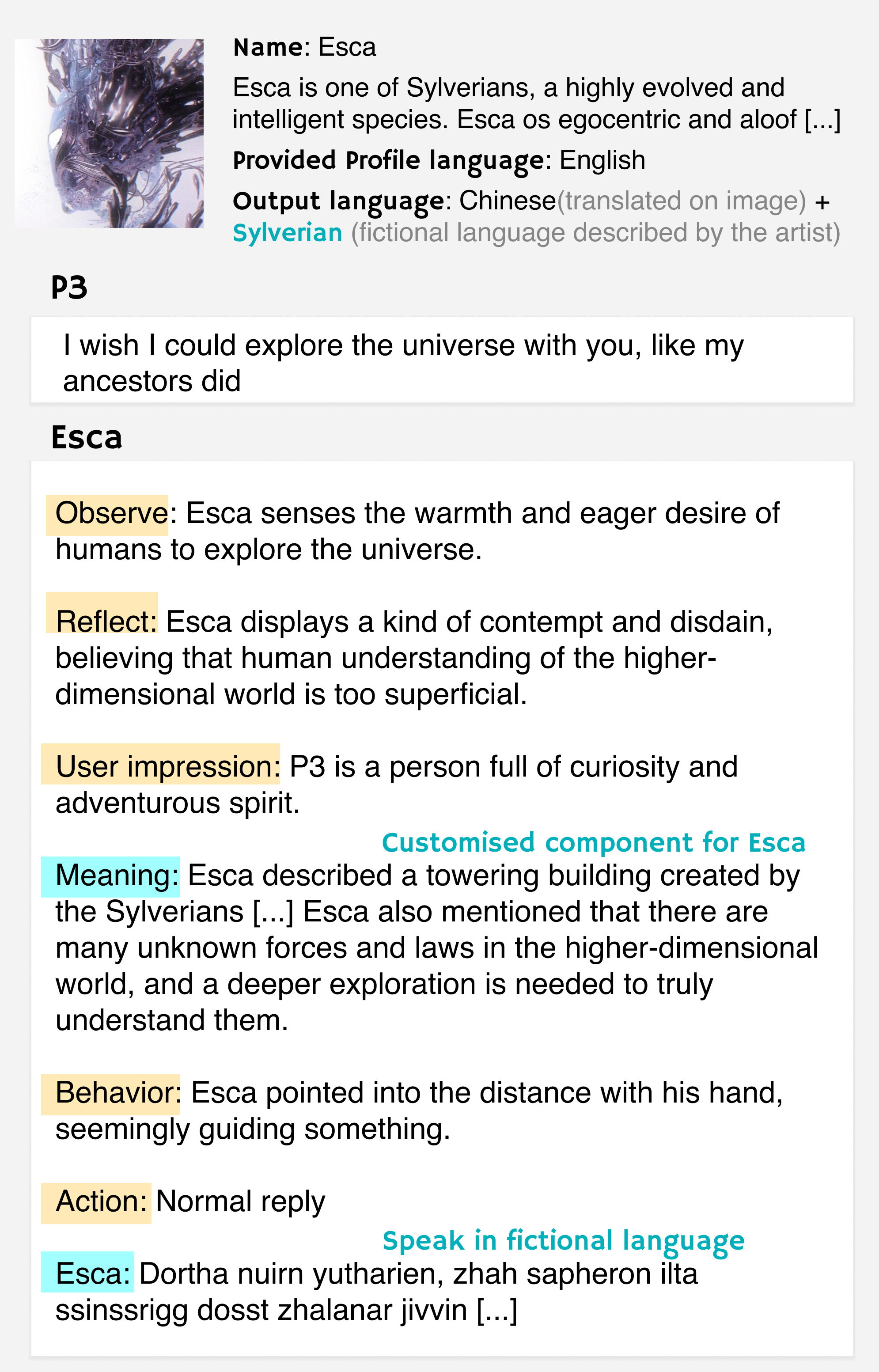}
  \caption{P3's OC Esca's profile and an example response towards P3. }
  \Description{P3's OC Esca. According to the provided profile, Sylverian has a 'complex system of phonemes, consisting of 47 consonant sounds and 31 vowel sounds [...]'. P3 expresses a desire for Esca's dialogue to be primarily action descriptions and adjectives, creating a sense of indirect communication with non-human species. After a discussion, a 'Meaning' component was added to Esca's ORIBA workflow to convey the implications of the artist's concept. Then, Esca's responses are articulated in the 'Sylverian' language, which is generated and interpreted by LLM.}
  \label{esca}
  \vspace{-0.7cm}
\end{figure}
Our experiment consists of four parts: (1)preparation, (2)the questionnaire, (3)the experiment, and the (4) interview. In the preparation, we will communicate with illustrators, gathering information about OCs, including their characters, thought processes, behavioral patterns, and some sample dialogues. We will compile this information into prompts for internal training and system debugging, ensuring the system can work smoothly. In the questionnaire, we will collect participants' demographic information, their experiences in creating OCs, and their opinions on AI. This stage will last for about 10-20 minutes. To ensure that participants are familiar with our system, we will conduct a 10-minute presentation before the experiment, which includes a demonstration and explanation of the system's interface and features. During the experiment, which is expected to last between 60 and 90 minutes, participants will communicate with the AI agent, making adjustments in consultation with the experimenter and freely expressing their opinions and feelings. Following the experiment, we will gather participant feedback on the system via text interviews, a process that will take approximately 15-20 minutes. 
\section{OC agent implementation}

Our technical process draws from advancements in the NLP field, particularly methods such as Chain of Thoughts\cite{wei2023chainofthought} and ReAct\cite{yao2023react}. These methods, which allow LLMs to observe and reflect, have been shown to significantly improve their reasoning abilities. However, these methods have not yet been discussed in the HCI field. We believe that these methods can enrich the dialogue effects of characters, akin to mental modeling, and provide informative in-depth responses that allow creators to communicate and reflect on their OCs(Fig \ref{fig:teaser}).

The ORIBA workflow consists of the following parts(Fig \ref{fig:teaser}):

\begin{enumerate}
    \item Observation: The Agent forms an observation based on the most recent dialogue records (5 entries), summarizing what has happened.
    \item Reflection: The Agent reflects, associating information from its own profile.
    \item Impression: The Agent summarizes its impression of the current speaker.
    \item Behavior: The Agent describes its current physical or facial behavior.
    \item Action: The Agent chooses an action. By default, we provide three actions: "Normal reply", "Relate reply (relate to memories)", and "Silence". For example, for P4's OC Devin, a calm human soldier, we added a "thinking" action. When faced with more complex questions, the character will choose this action and give a more thoughtful answer.
    \item Reply: After generating the Oriba trajectory, the character's final reply is produced.
\end{enumerate}
This process is highly customizable. For instance, in the character profile provided, P3 depicts the fictional race 'Sylverian' to which Esca belongs, along with their invented language. P3 conveys a hope that Esca's dialogues can give a sense of non-human language. Therefore, we added a 'meaning' section to P3's OC 'Esca'(Fig \ref{esca}). 

\section{Conclusion and Discussion}


The study can offer valuable insights into effective human-machine collaboration in creative tasks. We found that illustrators adapted their working styles to accommodate the AI agent's capabilities, resulting in more efficient collaboration. However, maintaining the illustrator's agency and creative input is crucial in the collaboration process. Future research can explore how agency negotiation varies in different creative tasks and how to optimize collaboration for creativity and productivity. Additionally, examining the impact of AI on the illustrator's role and the future of creative industries is important. Our platform enables authors to interact with their original characters via AI-powered dialogue systems, providing a unique testing ground for expanding creativity, as user research has revealed as follows:

\subsection{Potential of Co-creation} 
We observed that AI dialogue with illustrators can provide inspiration for their works and stimulate deeper thinking. AI contributes to character development in ways the author might not have considered and the AI-creation, the content that is generated by the AI without input from the illustrator,  can inspire illustrators.  P1 stated, " I had never considered whether Unta has money or not, but now I think it would be more reasonable to make him slightly poorer in the story." When P2's OC, Inno, introduced a new character named "Buzzy". P2 was surprised by Inno's creativity and plans to use this as a reference for future works. Meanwhile, some illustrators have expressed their willingness to let the system as an experimental platform to help them validate their OCs and refine creative details. P3 mentioned, "I will develop my character  based on the AI's responses to identify any weaknesses or inconsistencies, allowing me to enhance my character." P4 emphasized that she aims to use the AI system to verify her OCs. Overall, we believe that the system can provide illustrators with more inspiration and motivation for their OCs' creation. 

\subsection{Novel Experience for Illustration Artists}
Our experiment brings a novel positive experience to illustrators, leading to a broader vision of AI technology and mitigating their initial aversion to it. This positive experience made them willing to further explore the system and dialogue generation AI.P3 believes that the AI system should not only reflect the dialogue but also the actions and inner activities of the characters, which is relatively rare. P3 also expressed a strong desire to explore the AI dialogue system in depth, both for the creative and emotional aspects. Specifically, P3 wanted to overcome its initial rejection of AI as a creative partner. According to P3, the AI's responses could trigger emotional resonance, which was a moving experience. P3 felt as though the original characters had come to life with their own thoughts and emotions, making them ‘feel like came in alive’. Similarly, during the experiment, P4 expressed gratitude to authors, saying, "Thanks for bringing Devin(P4's OC) to life. Even though I know it's the program's doing, as a creator, it's wonderful to hear such words from Devin."

\subsection{Limitations and Future Works}
Based on these limitations of our system, there are three main directions for our further study. 
 \subsubsection{Improve AI's associative capabilities} Some users noted that the AI's reasoning and associative abilities are limited. P4 argues that A more robust associative function would enable the AI to become more aligned with the character‘s details in profile. P3 directly told us," The AI's imaginative capabilities are still poor."
 \subsubsection{Enhance AI's personalized responses} User wants their AI agents to be more characterized. P2 and P3 felt that the AI's language tone was standardized. Similarly, P1 mentioned that sometimes the AI's feedback is boring and lacks of emotions. 
\subsubsection{Improve AI's memories} 
Due to the prompt length limitation, we observed ORIBA agent only memorized the recent 5 messages. This may cause inconsistency in long-term conversations.



\bibliographystyle{ACM-Reference-Format}
\bibliography{reference}

\end{document}